\documentclass[prd,twocolumn,superscriptaddress,floatfix,amsmath,amssymb,amsfonts,longbibliography,nofootinbib]{revtex4-2}

\usepackage{float} 
\usepackage{comment}
\usepackage[normalem]{ulem}
\usepackage[english]{babel}
\usepackage{graphicx}
\usepackage{dcolumn}
\usepackage{bm}
\usepackage{blindtext}
\usepackage{verbatim}
\usepackage{mathrsfs}
\usepackage{amsmath}
\usepackage{cancel}
\usepackage{physics}
\usepackage{epstopdf}
\usepackage{mathtools}
\usepackage{tensor}
\usepackage{color}
\usepackage{xcolor}
\usepackage[usenames,dvipsnames]{pstricks}
\usepackage{epsfig}
\usepackage{pst-grad} 
\usepackage{pst-plot} 
\usepackage{hyperref}
\hypersetup{colorlinks,allcolors=black}
\usepackage{verbatim}
\usepackage{slashed}
\usepackage{dsfont}

\newcommand{\mf}{\mathsf}

\newcommand{\ii}{\mathrm{i}}

\newcommand{\M}{\mathcal{M}}

\newcommand{\tc}[1]{\textsc{#1}}

\allowdisplaybreaks[1] 

\begin{document}

\title{The effect of curvature on local observables in quantum field theory}

\author{Ahmed Shalabi}
\email{ashalabi@uwaterloo.ca}
\affiliation{Department of Physics and Astronomy, University of Waterloo, Waterloo, Ontario, N2L 3G1, Canada}

\author{Matheus H. Zambianco}
\email{mhzambia@uwaterloo.ca}
\affiliation{Department of Applied Mathematics, University of Waterloo, Waterloo, Ontario, N2L 3G1, Canada}
\affiliation{Perimeter Institute for Theoretical Physics, Waterloo, Ontario, N2L 2Y5, Canada}
\affiliation{Institute for Quantum Computing, University of Waterloo, Waterloo, Ontario, N2L 3G1, Canada}

\author{T. Rick Perche}
\email{trickperche@perimeterinstitute.ca}
\affiliation{Department of Applied Mathematics, University of Waterloo, Waterloo, Ontario, N2L 3G1, Canada}
\affiliation{Perimeter Institute for Theoretical Physics, Waterloo, Ontario, N2L 2Y5, Canada}
\affiliation{Institute for Quantum Computing, University of Waterloo, Waterloo, Ontario, N2L 3G1, Canada}

\begin{abstract}

  We compute the leading order corrections to the expected value of the squared field amplitude of a massless real scalar quantum field due to curvature in a localized region of spacetime. We use Riemann normal coordinates to define localized field operators in a curved spacetime that are analogous to their flat space counterparts, and the Hadamard condition to find the leading order curvature corrections to the field correlations. We then apply our results to particle detector models, quantifying the effect of spacetime curvature in localized field probes.
\end{abstract}

\maketitle

\section{Introduction}





The interplay between quantum theory and gravity remains a key challenge in physics. Despite significant progress, a complete theory of quantum gravity that seamlessly integrates these two foundational pillars is yet to be realized. Nevertheless, quantum field theory (QFT) in curved spacetimes provides a consistent approach to studying quantum systems under the influence of weak gravitational fields. This framework has proven invaluable for investigating phenomena such as particle creation in expanding universes~\cite{ParkerOG,ParkerIvanAgullo} and Hawking radiation~\cite{HawkingRadiation,Hawking1975,HawkingGibons}, offering insights into how quantum effects manifest due to curved spacetime backgrounds.






A relevant question in the context of QFT in curved spacetimes is how spacetime curvature influences the results of localized measurements of quantum fields. While the role of gravity in the local behaviour of correlations in quantum field theory is very well understood through the Hadamard condition~\cite{fullingHadamard,fullingHadamard2,kayWald,hadamardInArbitraryDim,fewsterNecessityHadamard}, the field correlations do not immediately translate to the expected values of field observables in a finite region of spacetime. Our goal in this paper is to explicitly connect the well-understood effect of curvature on the two-point functions of quantum fields to the physically accessible expected values of localized field operators.

To address this problem, we adopt the algebraic formulation of quantum field theory, which provides a versatile and rigorous description of localized observables in QFT. In this approach, a quantum field theory consists of an algebra of field operators, which are localized by compactly supported spacetime smearing functions. Moreover, the shape of these smearing functions can be directly linked to the localization of probes that interact with the field, and effectively implement measurements~\cite{fewster2,chicken,QFTPD}. In essence, the mathematically precise formulation of algebraic quantum field theory can also be seen as an operational formulation that explicitly identifies the observables that localized probes have access to.

Within this framework, we can draw an analogy between localized field observables for a real scalar quantum field in Minkowski spacetime and observables defined in a general curved spacetime. This comparison can be done by using the same shape for the localization regions in flat space and in curved space using Riemann normal coordinates, as those can be understood as a continuous deformation of inertial Minkowski coordinates naturally suited to describe the local effects of curvature. Together with curvature corrections given by the Hadamard condition, we identify the leading-order effects of curvature on the expected value of sufficiently localized observables of a real scalar quantum field. We then apply our results to particle detector models, such as the widely used Unruh-DeWitt detectors~\cite{Unruh1976,DeWitt} and compute the effect of curvature in the evolution of these local probes of quantum field theory.

This manuscript is organized as follows. In Section~\ref{sec:QFT} we briefly review the axiomatic formulation of quantum field theory and how to understand quantum fluctuations in terms of expected values of localized field observables. In Section~\ref{sec:curvature} we compute the effect of curvature on the expected value of the squared field amplitude localized in a Gaussian spacetime region. Then, we discuss how to operationally access these expected values utilizing local probes of quantum fields in Section~\ref{sec:detectors}. Finally, we discuss the conclusions of our work in Section~\ref{sec:conclusion}.

\section{quantum field theory in curved spacetimes}\label{sec:QFT}

In this section, we briefly review the algebraic formulation of the quantum field theory for a scalar field and set up the notations and conventions for the remainder of the manuscript. For a more complete description, we refer the reader to~\cite{Haag,araki1999mathematical,kasiaFewsterIntro,Khavkine_2015}.

A minimally coupled real Klein-Gordon scalar field $\phi$ in a globally hyperbolic spacetime $\mathcal{M}$ satisfies the following equation of motion
\begin{equation}
    P\phi = \nabla^{\mu} \nabla_{\mu}\phi = 0,
\end{equation}
where $\nabla_\mu$ denotes the covariant derivative arising from the Levi-Civita connection. The differential operator $P$ defines the retarded and advanced Green's functions $G_R(\mf x, \mf x')$ and $G_A(\mf x, \mf x')$ through the equation
\begin{equation}
\label{eq:G}
    \nabla^\mu \nabla_\mu G_{R/A}(\mf x, \mf x') = \delta(\mf x, \mf x'), 
\end{equation}
where $\delta(\mf x, \mf x')$ is the covariant Dirac delta distribution in spacetime~\cite{poisson}. Solutions to the homogeneous equation of motion can be obtained in terms of the causal propagator
\begin{equation}
    E(\mf x, \mf x') = G_R(\mf x, \mf x') - G_A(\mf x, \mf x').
\end{equation}
Indeed, for any real scalar function $\Lambda(\mf x)$, we have that 
\begin{equation}
    \phi(\mf x) = \int \dd V' E(\mf x, \mf x') \Lambda(\mf x')
\end{equation}
is a solution to the equation of motion, where $\dd V$ is the spacetime volume element. The causal propagator also plays a fundamental role in the commutation relations of the quantum field theory.

The quantum theory for a real scalar field can be built from a linear mapping $\hat{\phi}$ that acts in the space of smooth and compactly supported\footnote{Notice that although the standard formulation of AQFT utilizes compactly supported functions, more general spaces of test functions can also be considered, such as Schwartz.} functions on the globally hyperbolic spacetime $\mathcal{M}$. Each compactly supported function $\Lambda \in C_0^\infty(\Lambda)$ is then assigned to a symbol $\hat{\phi}(\Lambda)$. The elements $\hat{\phi}(\Lambda)$, together with an identity operator $\openone$ are then used to generate a $\ast$-algebra $\mathcal{A}(\mathcal{M})$ that satisfies the following conditions:
\vspace{2mm}

\noindent \textit{Hermiticity}: $(\hat{\phi}(\Lambda))^{\dagger} = \hat{\phi}(\Lambda^*)$, where $\dagger$ is the conjugation operation in $\mathcal{A}(\mathcal{M})$.
\vspace{2mm}

\noindent \textit{Field equation}: $\hat{\phi}(\Lambda) = 0$ for $\Lambda \in C_0^{\infty}(\mathcal{M})$.
\vspace{2mm}

\noindent \textit{Canonical commutation relations}: $$[\hat{\phi}(\Lambda_1),\hat{\phi}(\Lambda_2)] = \ii E(\Lambda_1,\Lambda_2), \,\text{ for }\, \Lambda_1,\Lambda_2 \in C_0^{\infty}(\mathcal{M}).$$

\noindent The $\ast$-algebra $\mathcal{A}(\mathcal{M})$ corresponds to the algebra of observables of the theory.

The algebra elements of the form $\hat{\phi}(\Lambda)$ can be understood as field operators smeared over the region supported by the function $\Lambda$. That is, one can formally write 
\begin{equation}
    \hat{\phi}(\Lambda) = \int \dd V \hat{\phi}(\mf x)\Lambda(\mf x).
\end{equation}
Intuitively, the smeared field operators correspond to the effective field operator that one has access to if one has access to the field in a spacetime region defined by the profile of $\Lambda(\mf x)$. Physically, the profile of $\Lambda(\mf x)$ is associated with the ``shape of the probe'' used to extract information about the field, as we will discuss in detail in Section~\ref{sec:detectors}.

In AQFT, quantum states are $\mathbb{C}$-linear functionals that map from the *-algebra $\mathcal{A}(\mathcal{M})$ to the complex numbers: $\omega:\mathcal{A}(\mathcal{M} )\to \mathbb{C}$ such that
\begin{align}
    &\omega(\openone) = 1, \nonumber \\
    &\omega(\hat{A}^{\dagger}\hat{A}) \geq 0 , \forall \hat{A} \in \mathcal{A}(\mathcal{M}).
\end{align}
In essence, a state is defined as a normalized positive linear functional that maps observables to their expected values:
\begin{equation}
    \langle\hat{A}\rangle_\omega \coloneqq \omega(\hat{A}).
\end{equation}
In particular, the n-point functions of the field are defined as the distributions
\begin{align}
    W(\Lambda_1,...,\Lambda_n) &\coloneqq \omega(\hat{\phi}(\Lambda_1)...\hat{\phi}(\Lambda_n)), 
\end{align}
which contain information about the field correlations between the regions defined by the profile of the functions $\Lambda_1,...,\Lambda_n$. A state is then fully determined by the set of all its n-point functions. 


A particularly relevant type of state in the context of quantum field theory are the so-called quasifree states~\cite{Haag,Khavkine_2015}, describing thermal states, most common choices of vacua and eigenstates of the number operator (when it can be defined). Quasifree states are zero-mean Gaussian states, so that the expected value of their odd-point functions vanish (in particular $\omega(\hat{\phi}(\Lambda)) = 0$) and their even-point functions are entirely determined by their two-point function
\begin{align}\label{eq:Wightman}
    W(\Lambda_1,\Lambda_2) &\coloneqq \omega(\hat{\phi}(\Lambda_1)\hat{\phi}(\Lambda_2)) \\
    &=  \int \dd V \dd V' W(\mf x, \mf x') \Lambda_1(\mf x) \Lambda_2(\mf x'),\nonumber
\end{align}
where the other even-point functions can be obtained through Wick's theorem~\cite{Wicked}. The integral kernel $W(\mf x, \mf x')$ in Eq.~\eqref{eq:Wightman} is often referred to as the \textit{Wightman function}. The Wightman function characterizes important aspects of a quantum field theory including field correlations, the UV behaviour of the theory~\cite{kayWald,fullingHadamard,fullingHadamard2}, as well as complete information about the background geometry of spacetime~\cite{achim,achim2,geometry}. 



Interestingly, if one has access to the quantum field in a quasifree state in a region defined by the spacetime profile of a function $\Lambda(\mf x)$, the expected value of the corresponding localized field operator $\hat{\phi}(\Lambda)$ is zero. Intuitively, this implies that over many ``measurements'', the results obtained for the field amplitude average out to zero. The ``measurement results'' would then be entirely determined by the variance
\begin{equation}
    \langle \Delta \hat{\phi}(\Lambda)\rangle_\omega = \langle\hat{\phi}(\Lambda)^2\rangle_\omega - \langle\hat{\phi}(\Lambda)\rangle_\omega^2 = \langle\hat{\phi}(\Lambda)^2\rangle_\omega,
\end{equation}
which also corresponds to the fluctuations of the field amplitude in the state $\omega$ in the spacetime region defined by the profile of $\Lambda(\mf x)$. The operator $\hat{\phi}(\Lambda)^2$ is indeed the simplest non-trivial observable that can be accessed in the region defined by $\Lambda(\mf x)$. Our goal in this paper will be to compute the leading order effect of curvature to the expected value $\langle \hat{\phi}(\Lambda)^2\rangle_\omega$ when $\Lambda(\mf x)$ is a sufficiently small spacetime region. 





\subsection*{Massless scalar field in Minkowski spacetime}

To be able to find the effect of curvature in the expected value $\langle \hat{\phi}(\Lambda)^2\rangle_\omega$, we first review explicit results for the Minkowski vacuum. Let $\hat{\phi}_0(\mf x)$ denote a real scalar field in Minkowski spacetime. The symmetries of the spacetime single out a unique state $\omega_0$ that is invariant under Poincar\'e transformations. As with any other quasifree state, the Minkowski vacuum is fully determined by its Wightman function $W_0(\mf x, \mf x') = \omega_0(\hat{\phi}_0(\mf x) \hat{\phi}_0(\mf x'))$. In inertial coordinates $\mf x = (t,\bm x)$, the Wightman function can be written as
\begin{equation}\label{eq:WMink}
    W_0(\mf x,\mf x') = \lim_{\epsilon\to 0^+}\frac{1}{4\pi^2(-(t-t' - \ii \epsilon)^2 + (\bm x - \bm x')^2)}.
\end{equation}
The necessity to add the regulator $\ii \epsilon$ is a general feature of two-point functions in quantum field theory, as the Wightman function is in general divergent in the limit $\mf x' \to \mf x$. This divergence does not change the fact that $W(\Lambda_1,\Lambda_2)$ is finite for any smooth and sufficiently fast decaying functions $\Lambda_1$ and $\Lambda_2$. As a rule of thumb, we will omit the regulator $\ii \epsilon$ throughout the manuscript, taking it into account when necessary for integration.



The Minkowski vacuum Wightman function can also be written in terms of a plane wave expansion:
\begin{equation}\label{eq:Wmomentum}
    W_0(\mf x, \mf x') = \frac{1}{(2 \pi)^3}\int \frac{\dd^3 \bm k}{2|\bm k|} e^{\ii \mf k \cdot (\mf x - \mf x')},
\end{equation}
where $\mf k = (|\bm k|,\bm k)$ and $\mf k \cdot \mf x = \eta_{\mu\nu} k^\mu x^\nu$, with \mbox{$\eta_{\mu\nu} = \text{diag}(-1,1,1,1)$} being the Minkowski metric. Notice that Eq.~\eqref{eq:Wmomentum} is valid in distributional form, and technically also requires regularization with $t-t' \mapsto t-t' - \ii \epsilon$, which selects the correct poles of the integrand in the complex plane~\cite{birrell_davies}.


Having the global expression for the Wightman function of the Minkowski vacuum then allows us to write $\langle\hat{\phi}_0(\Lambda)^2\rangle$ explicitly as
\begin{equation}
    \langle\hat{\phi}_0(\Lambda)^2\rangle = \int \dd^4\mf x \dd^4 \mf x' W_0( \mf x, \mf x')\Lambda(\mf x) \Lambda(\mf x'),
\end{equation}
where the function $\Lambda(\mf x)$ defined the region of spacetime where one has access to the field. A particularly versatile example is when we consider the expected value of the field observable $\langle\hat{\phi}(\Lambda)^2\rangle$ in a region defined by the Gaussian test function 
\begin{equation}
\label{eq:LambdaGaussian}
    \Lambda(\mf x) = \frac{e^{-\frac{t^2}{2T^2}}}{\sqrt{2\pi}T}\frac{e^{-\frac{\abs{\bm x}^2}{2\sigma^2}}}{(2\pi \sigma^2)^{3/2}},
\end{equation}
where $T$ defines the effective temporal extension of $\Lambda(\mf x)$ and $\sigma$ defines its spatial localization. In the limit of $\sigma,T\to 0$, the function $\Lambda(\mf x)$ yields a Dirac delta, centered at the origin\footnote{Notice that due to Poincar\'e invariance of the Minkowski vacuum, a spacetime translation of the form $\Lambda(\mf x)\mapsto \Lambda(\mf x - \mf x_0)$ does not affect the expected value of the observable $\langle\hat{\phi}_0(\Lambda)^2\rangle$.}. For sufficiently small $\sigma$ and $T$, the observable $\hat{\phi}_0(\Lambda)^2$ then corresponds to the variance of the field amplitude at the origin, when it is probed in a region of time and space extensions defined by $T$ and $\sigma$, respectively.

An advantage of considering $\Lambda(\mf x)$ as given in Eq.~\eqref{eq:Wmomentum} is that we can compute $W_0(\Lambda,\Lambda)$ explicitly. Indeed, using the plane wave expansion on Eq.~\eqref{eq:LambdaGaussian}, we can write $\langle\hat{\phi}(\Lambda)^2\rangle$ in terms of an integral in momentum space, namely
\begin{equation}
    \langle\hat{\phi}_0(\Lambda)^2\rangle = \frac{1}{(2\pi)^3}\int \frac{\dd^3 \bm k}{2|\bm k|} |\tilde{\Lambda}(|\bm k|,\bm k)|^2,
    \label{eq:momentum_space}
\end{equation}
where we define the spacetime Fourier transform
\begin{equation}
    \tilde{\Lambda}(|\bm k|,\bm k) \coloneqq \int \dd^4 \mf x \Lambda(\mf x) e^{\ii \mf k \cdot \mf x}.
    \label{eq:fourier_ladrao}
\end{equation}
For the specific choice of $\Lambda(\mf x)$ of Eq.~\eqref{eq:LambdaGaussian}, we obtain
\begin{equation}
    \langle\hat{\phi}_0(\Lambda)^2\rangle = \frac{1}{8\pi^2(\sigma^2 + T^2)}.
    \label{eq:nice}
\end{equation}
The quantity above quantifies the fluctuations of the Minkowski vacuum when it is sampled in a Gaussian spacetime region of time duration $T$ and spatial duration $\sigma$. As expected, this quantity is divergent in the limit \mbox{$\sigma,T\to 0$} due to the standard UV divergences in quantum field theory.

\section{The effect of curvature on field observables}\label{sec:curvature}

In this section, we will compute the effect of curvature on the expected value of an operator of the form $\hat{\phi}(\Lambda)^2$, where $\Lambda(\mf x)$ is a spacetime function, with support centered at an event $\mf z$. We consider the quantum field theory for a massless scalar field $\hat{\phi}(\mf x)$ in a general globally hyperbolic spacetime $\mathcal{M}$, according to the construction reviewed in Section~\ref{sec:QFT}. We assume the extension both in space and time of the function $\Lambda(\mf x)$ to be characterized by a parameter $\ell$ with dimensions of length. The expected value of $\hat{\phi}(\Lambda)^2$ on a state $\omega$ can be written as
\begin{equation}
    \langle\hat{\phi}(\Lambda)^2\rangle_\omega = \int \dd^4 \mf x \dd^4 \mf x' \sqrt{-g} \sqrt{-g'}W(\mf x, \mf x') \Lambda(\mf x) \Lambda(\mf x'),\label{eq:intphi2}
\end{equation}
where we wrote the volume form explicitly in terms of the metric determinant $\sqrt{-g}$. 

In principle, it is not trivial to compare quantum fluctuations in a curved background spacetime to the ones in Minkowski spacetime. This is mainly because of two reasons: 1) The regions where the field is probed in Minkowski spacetime and in general spacetime cannot be directly compared, as they are defined by functions in different manifolds and 2) The quantum field theories are in principle entirely different, defined on different manifolds with different equations of motion. Nevertheless, both of these issues can be bypassed by considering that the spacetime region where $\Lambda(\mf x)$ is supported (controlled by the parameter $\ell$) is sufficiently small compared to the radius of curvature of the spacetime. 

The issue of comparing the spacetime smearing functions in Minkowski spacetime and in the general spacetime $\mathcal{M}$ in sufficiently small regions of spacetime can be addressed by employing Riemann normal coordinates centered at the event $\mf z$ at the center of the support of $\Lambda(\mf x)$. In essence, Riemann normal coordinates are an analogue to inertial coordinates in Minkowski spacetime, where the distance between an event $\mf x$ with coordinates $x^a$ and the center of Riemann normal coordinates, $\mf z$, is given by $\eta_{ab} x^ax^b$~\cite{poisson}. That is, if the function $\Lambda_0(t,\bm x)$ written in inertial coordinates in Minkowski spacetime has its maximum at the origin, we can ``import'' this function to $\mathcal{M}$ by considering the function $\Lambda(\mf x)$ defined by $\Lambda(x^a) = \Lambda_0(x^0,x^1,x^2,x^3)$ in Riemann normal coordinates\footnote{Notice that the Riemann normal coordinates centered at an event $\mf z$ are only defined in the normal neighbourhood of $\mf z$, so that $\Lambda(x^a)$ is technically only defined in this region. If $\Lambda_0(t,\bm x)$ is compactly supported within the domain of the Riemann normal coordinates, then it can be extended to a compaclty supported function in $\M$. If the function $\Lambda_0(\mf x)$ does not have compact support (such as the example in Eq.~\eqref{eq:LambdaGaussian}), one effectively technically needs to add a hard-cutoff to its tails to define $\Lambda(\mf x)$ as a compactly supported function in $\mathcal{M}$.}. This ensures that the spacetime support of $\Lambda(\mf x)$ behaves in the same way in which the support of $\Lambda_0(\mf x)$ behaves in Minkowski spacetime, in the following sense. Given a region $U$ in Minkowski spacetime, one can define its maximal geodesic size as
\begin{equation}
    \ell_U \equiv \sup_{\mf x, \mf x' \in U} \sqrt{2|\sigma(\mf x, \mf x')|},
\end{equation}
corresponding to the maximal spacetime separation between any two events in $U$. Then, if $\Lambda_0(\mf x)$ is supported in a region that contains the origin, with maximal geodesic size $\ell$ in Minkowski, $\Lambda(\mf x)$ will be compactly supported in a subregion of $\M$
whose size is approximately $\ell$, provided that $\ell$ is sufficiently small. An example is if the support of $\Lambda_0(\mf x)$ is contained in a \textit{Euclidean} ball of radius $\ell/2$ in Minkowski spacetime.

We can compare the Wightman function in a given curved spacetime with that of the Minkowski vacuum by assuming that the state $\omega$ is quasifree and satisfies the so-called Hadamard condition. The Hadamard condition is a statement about the local behaviour of the Wightman function and ensures the renormalization of the stress-energy tensor of a QFT~\cite{fullingHadamard,fullingHadamard2,kayWald,hadamardInArbitraryDim,fewsterNecessityHadamard}. In essence, it states that when $\mf x'$ is in a normal neighbourhood of $\mf x$ (which we assume to be overlapping with the normal neighbourhood of $\mf z$), the Wightman function can be put in the form
\begin{align}
    W(\mf x,\mf x') = \frac{\Delta^{1/2}( \mf x,\mf x')}{8\pi^2 \sigma( \mf x, \mf x')} + v( \mf x,\mf x') \text{ln}\left(\frac{\sigma(\mf x, \mf x')}{\ell_0^2}\right)+w(\mf x,\mf x'),\label{eq:Hadamard}
\end{align}
\noindent where $\sigma(\mf x, \mf x')$ is Synge's world function~\cite{Synge,poisson} (corresponding to one half the squared geodesic distance between $\mf x$ and $\mf x'$), $\ell_0$ is an arbitrary parameter with units of length~\footnote{Notice that the specific value of the parameter $\ell_0$ does not affect the divergence structure of the Wightman function, as it can be incorporated into the state-dependent term $w(\mf x, \mf x')$:
$$\ell_0\mapsto \ell_0' \quad\Rightarrow\quad w(\mf x, \mf x') \mapsto w(\mf x, \mf x') - 2\text{ln}(\ell_0'/\ell_0)v(\mf x, \mf x').$$}, $\Delta(\mf x, \mf x')$ is the Van-Vleck determinant~\cite{Van_Vleck1928}:
\begin{equation}
    \Delta\left(\mf x, \mf x^{\prime}\right)=-\frac{\operatorname{det}\left(-\sigma_{\alpha \beta^{\prime}}\left(\mf x, \mf x^{\prime}\right)\right)}{\sqrt{-g} \sqrt{-g^{\prime}}},
\end{equation}
where we use the standard notation for the derivatives of Synge's world function, $\sigma_{\alpha} =\partial_{\alpha} \sigma$. Finally, $v(\mf x, \mf x')$, $w(\mf x, \mf x')$ are functions that admit regular expansions as a function of $\sigma(\mf x, \mf x')$,
\begin{align}
    v(\mf x, \mf x') &= \sum_{n=0}^\infty v_n(\mf x, \mf x')(\sigma(\mf x, \mf x'))^n,\label{eq:vExp}\\
    w(\mf x, \mf x') &= \sum_{n=0}^\infty w_n(\mf x,\mf x')(\sigma(\mf x, \mf x'))^n.
\end{align}
Importantly, $v(\mf x, \mf x')$ is entirely determined by the geometry of the background spacetime, and $w(\mf x, \mf x')$ encodes the information about the state $\omega$. Notice that our assumption that $\ell$ characterizes the effective size of the support of $\Lambda(\mf x)$ translates to the assumption that $\sigma(\mf x,\mf z),\sigma(\mf x',\mf z),\sigma(\mf x,\mf x')$ are all of order $\mathcal{O}(\ell^2)$ within the support of $\Lambda(\mf x)$.

To find the effect of curvature on an expected value of the form $\langle\hat{\phi}(\Lambda)^2\rangle_\omega$, we will first rewrite the Wightman function $W(\mf x, \mf x')$ in terms of an analogue expression for the Wightman function in Minkowski spacetime, namely
\begin{equation}
    W_\sigma(\mf x, \mf x') = \frac{1}{8\pi^2 \sigma(\mf x, \mf x')},
\end{equation}
added to corrections of higher order in $\sigma(\mf x, 
\mf x')$. This will allow us to expand the integral of Eq.~\eqref{eq:intphi2} in terms of $\ell$ using that $|\sigma|\lesssim \ell^2$ when smeared by $\Lambda(\mf x)$. Factoring $W_0(\mf x, \mf x')$, we find
\begin{align}
    W(\mf x,\mf x') = \frac{\Delta^{1/2}}{8\pi^2\sigma}  +  v_0&\text{ln}(\sigma/\ell_0)+ w_0 + \mathcal{O}(\sigma\,\text{ln}\sigma).\label{Wexpsigma}
\end{align}
We can now expand $\Delta(\mf x, \mf x')$, $v(\mf x, \mf x')$, and $w(\mf x, \mf x')$ in terms of Synge's world function. The Van-Vleck determinant
admits the expansion~\cite{poisson} 
\begin{equation}
    \Delta(\mf x,\mf x') = 1 + \frac{1}{6}R_{\alpha \beta}(\mf x) \sigma^\alpha(\mf x,\mf x') \sigma^\beta(\mf x,\mf x') + \mathcal{O}(\sigma^2).
\end{equation}
The vector $\sigma^\alpha$ also acts as an effective separation vector between the events $\mf x$ and $\mf x'$, corresponding to the initial vector of a geodesic that starts at $\mf x$ and reaches $\mf x'$~\cite{poisson}. For a non-conformally coupled Klein-Gordon equation, the coefficient $v_0(\mf x, \mf x')$ in Eq.~\eqref{eq:vExp} can also be expanded as~\cite{DeWittExpansion}
\begin{equation}
    v_0(\mf x, \mf x') = \frac{R(\mf x)}{12} + \mathcal{O}(\sigma).
\end{equation}
Combining these results, 
we can recast the Wightman function as
\begin{align}
    W(\mf x,\mf x') =  &\, W_\sigma(\mf x,\mf x')\left(1 + \frac{1}{12}R_{\alpha\beta}(\mf x)\sigma^\alpha \sigma^\beta\right) \label{WW0partial}
    \\
    &+\frac{1}{12}R(\mf x)\text{ln}\left(\sigma/\ell_0^2\right)+w_{0}(\mf x, \mf x') + \mathcal{O}(\sigma\ln\sigma).\nonumber
\end{align}
Our goal is now to write the integral that defines $\langle\hat{\phi}(\Lambda)\rangle_\omega$ in Eq.~\eqref{eq:intphi2} in Riemann normal coordinates (RNC), which will also allow us to perform an expansion in $\ell$, that is, an expansion in the effective size of the region that localizes the observable $\hat{\phi}(\Lambda)^2$. Essentially, for $\mf x, \mf x'$ within this region, we have $\sigma(\mf x, \mf x') = \mathcal{O}(\ell^2)$. The Riemann normal coordinates centered at $\mf z$ consist of a local diffeomorphism that effectively maps coordinates in the tangent space of $\mf z$ to its normal neighbourhood. As a consequence, given an orthonormal basis $\{e_\mu\}$ of $T_{\mf z}\mathcal{M}$, we can write the Riemann normal coordinates of an event $\mf x$ in the normal neighbourhood of  $\mf z$ as $x^a = \sigma^\mu(\mf z, \mf x)(e_{\mu})^a$. That is, in Riemann normal coordinates, expressions of the form $\sigma^a(\mf z, \mf x)$ directly correspond to the coordinates of the event $\mf x$. 
Furthermore, we can expand  $\sigma(\mf x, \mf x')$ in Riemann normal coordinates centered at $\mf z$ to leading order in $\ell$ as
\begin{equation}
    \sigma(\mf x, \mf x') \approx \frac{1}{2}\eta_{ab}(x-x')^a(x-x')^b + \frac{1}{6} R_{acbd}(\mf z) x^a x^b x'{}^c x'{}^d.
     \label{eq:camelsing}
\end{equation}
We refer the reader to Appendix~\ref{app:sing} for a derivation of the expression above. From this expression, it is straightforward to determine the expansions of the derivatives $\sigma_{a'}(\mf x, \mf x') \equiv \partial_{a'} \sigma(\mf x, \mf x')$ and $\sigma_{a}(\mf x, \mf x') \equiv \partial_{a} \sigma(\mf x, \mf x')$, namely
\begin{align}
    &\sigma^{a'}(\mf x, \mf x') \approx (x-x')^a + \frac{1}{3} R_{dcb}{}^a(\mf z) x^d x^b x'{}^d,\nonumber\\
    &\sigma^{a}(\mf x, \mf x')\approx (x-x')^a + \frac{1}{3} R^{a}{}_{cbd}{}(\mf z) x^b x'{}^c x'{}^d.
\end{align}


Combining the results above with the expansion of Eq.~\eqref{WW0partial} we can write the leading order behaviour of the Wightman function in a neighbourhood of the event $\mf z$ as
\begin{align}
    &W(\mf x,\mf x') \approx W_\sigma(\mf x,\mf x')\left(1 + \frac{1}{12}R_{ab}(\mf z)(x-x')^a(x-x')^b \right)\label{eq:WW0} \nonumber\\
    &+\frac{1}{12}R(\mf z)\text{ln}\left(\frac{(x-x')^2}{2\ell_0^2}\right)+w_0(\mf x,\mf x') + \mathcal{O}(\sigma\ln\sigma),
\end{align}
where we denote $(x-x')^2 = \eta_{ab} (x-x')^a (x-x')^b$.
The last step in our expansion of the Wightman function will be to expand $\sigma(\mf x, \mf x')$ in $W_\sigma(\mf x, \mf x')$ in Riemann normal coordinates centered at $\mf z$. We have
\begin{align}
    &W_\sigma(\mf x, \mf x') = \frac{1}{8\pi^2 \sigma(\mf x, \mf x')}\label{eq:Matheus}\\
    &= W_0(\mf x, \mf x')\left(1  - \frac{4\pi^2}{3}R_{acbd}x^a x^b x'{}^c x'{}^d W_0(\mf x, \mf x')\right) + \mathcal{O}(\sigma^2)\nonumber,
\end{align}
where
\begin{equation}
    W_0(\mf x, \mf x') = \frac{1}{4\pi^2 \eta_{ab}(x-x')^a(x-x')^b}.
\end{equation}
Notice that the correction term proportional to $x^a x^b x'{}^c x'{}^d W_0(\mf x, \mf x')$ is of order $\ell^2$.

The last step to fully write the expected value $\langle\hat{\phi}(\Lambda)^2\rangle_\omega$ in Riemann normal coordinates centered at $\mf z$ is to expand the metric determinant $\sqrt{-g}$ in Riemann normal coordinates centered at $\mf z$. The expansion reads~\cite{poisson}
\begin{equation}\label{eq:sqrtg}
    \sqrt{-g(\mf x)} =\frac{1}{\Delta(\mf z, \mf x)} =  1 - \frac{1}{6}R_{a b}(\mf z) \sigma^a(\mf z,\mf x) \sigma^b(\mf z,\mf x) + \mathcal{O}(\ell^3).
\end{equation}
Using Eqs.~\eqref{eq:WW0} and~\eqref{eq:sqrtg} in the integral expression of Eq.~\eqref{eq:intphi2}, we obtain
\begin{align}\label{eq:main}
    \langle\hat{\phi}(\Lambda)^2\rangle_\omega = & \langle\hat{\phi}_0 (\Lambda)^2\rangle -\frac{4 \pi^2}{3} R_{abcd}\mathcal{L}^{abcd}-\frac{1}{12} R_{a b} \mathcal{L}^{a b} \nonumber\\ & + \frac{1}{12} R \,\mathcal{P}_{\ln} + \omega_{\Lambda}
    + \mathcal{O}(\ell^2\ln\ell),
\end{align}
where
\begin{align}
&\langle\hat{\phi}_0(\Lambda)^2\rangle = \int\dd^4 \mf x \dd^4 \mf x' \Lambda(\mf x) \Lambda(\mf x') W_0(\mf x,\mf x'),\nonumber\\
     &\mathcal{L}^{abcd} =\int\dd^4 \mf x \dd^4 \mf x' \Lambda(\mf x) \Lambda(\mf x') W_0^2(\mf x,\mf x')x^{a}x^{d}{x'}^{b}{x'}^{c}\!,\nonumber \\
    &\mathcal{L}^{ab} =\int\dd^4 \mf x \dd^4 \mf x' \Lambda(\mf x) \Lambda(\mf x') W_0(\mf x,\mf x')(x+x')^a(x+x')^b\!,\nonumber\\
    &\mathcal{P}_{\ln} = \int\dd^4 \mf x \dd^4 \mf x' \Lambda(\mf x) \Lambda(\mf x')\text{ln}\left(\frac{(x-x')^2}{\ell_0^2}\right),\nonumber\\
    &\omega_{\Lambda}= \int\dd^4 \mf x \dd^4 \mf x' \Lambda(\mf x) \Lambda(\mf x')w_0(\mf x, \mf x').\label{Ls}
\end{align}
Notice that the term $\langle\hat{\phi}_0(\Lambda)^2\rangle$ corresponds to the exact same expression of the squared field amplitude in the Minkowski vacuum (see Eq.~\eqref{eq:WMink}). 
The remaining terms in Eq.~\eqref{eq:main} represent corrections to the squared field amplitude, up to order $\mathcal{O}(\ell^2\ln\ell)$. The coefficients $\mathcal{L}^{abcd}$, $\mathcal{L}^{ab}$, and $\mathcal{P}_{\ln}$ are associated with corrections to the expected value due to the curvature of spacetime, whereas the term $\omega_{\Lambda}$ is associated to the local effect of the field state in the expected value of the squared field amplitude over the region defined by the profile $\Lambda(\mf x)$. Notice that one could also perform a multipole expansion of $w_0(\mf x, \mf x')$, obtaining a series in $\ell$.

We can concretely evaluate Eq. \eqref{eq:main} using the Gaussian spacetime smearing of Eq. \eqref{eq:LambdaGaussian}, written in Riemann normal coordinates. In Appendix~\ref{app:L-terms} we compute the general expression of Eq.~\eqref{eq:main}, where the temporal and spatial profiles $T$ and $\sigma$ couple to different components of the curvature tensors. For simplicity, here we assume that the temporal and spatial supports of the Gaussian are the same ($T = \sigma = \ell$). The corrections due to the Riemann and Ricci tensors then read:
\begin{equation}
       -\frac{4 \pi^2}{3}R_{abcd}\mathcal{L}^{abcd}-\frac{1}{12} R_{a b} \mathcal{L}^{a b}  = 
       -\frac{5R + 3 R_{00}}{576 \pi^2}.
\end{equation}

Picking $\ell_0 = \ell$, we can also numerically evaluate $\mathcal{P}_{\ln}$ to get \mbox{$\mathcal{P}_{\ln} \approx -0.84961$}. Notice that other choices of $\ell_0$ would simply be absorbed by the state dependent term $\omega_0$. We find the leading order corrections to the expected value of the squared field amplitude of a massless field in a region of size $\ell$ to be given by:
\begin{align}
    \langle\hat{\phi}(\Lambda)^2\rangle_\omega = \frac{1}{16\pi^2\ell^2}-\frac{5R + 3 R_{00}}{576 \pi^2} + &\frac{1}{12}R\, \mathcal{P}_{\ln} + \omega_\Lambda \nonumber\\& + \mathcal{O}(\ell^2\ln\ell). 
\end{align}
Notice that the term $\omega_\Lambda$ is not entirely determined by the geometry of spacetime, and explicitly depends on the local properties of the field state being probed. To leading order in the region size $\ell$, the field state contributes only with a constant correction. Also, notice that the expression for $\langle\hat{\phi}(\Lambda)^2\rangle_\omega$ explicitly depends on the temporal direction chosen for the Riemann normal coordinates. This is not entirely unexpected, as even in Minkowski spacetime the very natural localizing function of Eq.~\eqref{eq:LambdaGaussian} is not Lorentz invariant.

\section{Effective probes of quantum fields}\label{sec:detectors}

In this section we will discuss how to physically access the smeared field operators of the form $\hat{\phi}(\Lambda)$, utilizing local probes in quantum field theory. As we previously mentioned, the spacetime smearing functions $\Lambda(\mf x)$ that define localized field observables can be associated with the localization of probes that couple to the field. These probes are usually refereed to as particle detector models~\cite{Unruh1976,DeWitt,Unruh-Wald}, and they have been extensively used to provide an operational perspective to localized operations in the context of QFT. In particular, particle detectors have been used in the context of numerous relativistic quantum information protocols, such as entanglement harvesting~\cite{Valentini1991,Reznik1,reznik2,Salton:2014jaa,Pozas-Kerstjens:2015,Pozas2016,bandlimitedHarv2020,ampEntBH2020,threeHarvesting2022,ericksonNew,carol,boris,MatheusAdam1}, quantum energy teleportation~\cite{teleportation2014,teleportation,nichoTeleport,teleportExperiment}, quantum collect calling~\cite{Jonsson2,collectCalling,PRLHyugens2015,Simidzija_2020}, among others~\cite{KojiCapacity,KojiEntTeleport}. 

The simplest and most often utilized particle detector model is the two-level Unruh-DeWitt~\cite{Unruh1976,DeWitt} detector. It consists of a qubit that undergoes a timelike trajectory $\mf z(\tau)$ and interacts linearly with a quantum field $\hat{\phi}(\mf x)$. The detector's free dynamics is prescribed by the Hamiltonian
\begin{equation}
    \hat{H}_\tc{d} = \Omega \hat{\sigma}^+\hat{\sigma}^-,
\end{equation}
which generates time evolution with respect to the time parameter $\tau$. $\hat{\sigma}^\pm$ are the $\mathfrak{su}(2)$ raising an lowering operators and $\Omega$ is the energy gap of the qubit system. The eigenstates of the Hamiltonian then define the ground and excited state of the system, $\{\ket{g},\ket{e}\}$ by $\hat{H}_\tc{d}\ket{g} = 0$ and \mbox{$\hat{H}_\tc{d}\ket{e} = \Omega \ket{e}$}. Using these states, we can write $\hat{\sigma}^{+} = | e\rangle \langle g| $ and $\hat{\sigma}^{-} = | g \rangle \langle e|$.

The interaction with the quantum field is prescribed by the interaction Hamiltonian density~\cite{us,us2}
\begin{equation}
\label{h_I}
    \hat{h}_I(\mf x) = \lambda \Lambda(\mf x) \hat{\mu}(\tau) \hat{\phi}(\mf x), 
\end{equation}
where $\tau$ denotes the Fermi normal coordinate time associated to the trajectory $\mf z(\tau)$~\cite{poisson,generalPD}, $\hat{\mu}(\tau) = e^{\ii \Omega \tau}\hat{\sigma}^+ +e^{\ii \Omega \tau}\hat{\sigma}^-$ is the time evolved monopole operator of the detector in the interaction picture, $\lambda$ is a coupling constant, and $\Lambda(\mf x)$ is the so-called spacetime smearing function, supported around the trajectory $\mf z(\tau)$. $\Lambda(\mf x)$ defines the profile of the interaction in both space and time, and as we will see, it also defines the localization of the smeared field operators that the detector can access. 

To see how a two-level UDW detector obtains information about field operators localized by $\Lambda(\mf x)$, we consider the process where the detector starts in its ground state and the quantum field in a quasifree state $\omega$ before the interaction (in the causal past of the support of $\Lambda(\mf x)$). One can obtain the final state of the detector by applying the time evolution operator
\begin{equation}
    \hat{U}_I = \mathcal{T}\exp(-\ii \int \dd V \hat{h}_I(\mf x))
\end{equation}
to the detector-field initial state and tracing over the field's degrees of freedom (for the explicit calculation we refer the reader to~\cite{Pozas-Kerstjens:2015,us2,ericksonNew}). The leading order excitation probability of the detector then reads
\begin{align}\label{eq:L}
    \mathcal{L} &= \lambda^2 \omega(\hat{\phi}(\Lambda^-)\hat{\phi}(\Lambda^+))\\
    &= \lambda^2 \int \dd V \dd V' \Lambda(\mf x) \Lambda(\mf x') e^{- \ii \Omega(\tau - \tau')}W(\mf x, \mf x'),\nonumber
\end{align}
where $\Lambda^\pm(\mf x) = e^{\pm \ii \Omega \tau} \Lambda(\mf x)$ and $W(\mf x, \mf x')$ is the Wightman function defined by the initial field state $\omega$. It is then clear that the detector obtains information about field operators smeared by the function $\Lambda(\mf x)$, with the addition of the phase terms $e^{\pm\ii \Omega \tau}$. The phase terms make it so that the integral of Eq.~\eqref{eq:L} samples frequencies of the correlation function that effectively resonate with the detector's energy gap.

Also notice that if $\Omega = 0$, $\Lambda^+(\mf x) = \Lambda^-(\mf x) = \Lambda(\mf x)$, and Eq.~\eqref{eq:L} becomes proportional to $\langle\hat{\phi}(\Lambda)\rangle_\omega$. However, in this case $\mathcal{L}$ cannot be interpreted as an excitation probability, as the detector does not have an energy gap. Gapless detectors have been studied in detail~\cite{Landulfo,JormaPozas,analytical}, partially because their interaction with the field can be solved non-perturbatively, but also for their ability to describe degenerate subsystems and UDW detectors in the limit where all other relevant parameters are significantly larger than $\Omega$~\footnote{Specifically, for all parameters with units of length, $\ell_i$, we must have $\Omega \ell_i\ll 1$. In our case, the relevant parameters are $T$ and $\sigma$.}. Setting $\Omega = 0$, the interaction Hamiltonian \eqref{h_I} reduces to 
\begin{equation}
    \hat{h}_I(\mf x) = \lambda \Lambda(\mf x) \hat{\mu} \hat{\phi}(\mf x)
\end{equation}
with $\hat{\mu} = \hat{\sigma}^+ + \hat{\sigma}^-$. The interaction Hamiltonian density in the expression above satisfies \mbox{$[[\hat{h}_I(\mf x),\hat{h}_I(\mf x')],\hat{h}_I(\mf x'')] = 0$}. This is the key fact that allows us to solve for the dynamics of this model non-perturbatively using the Magnus expansion~\cite{magnus,magnusReview}. The unitary time evolution operator for this gapless detector model can be written as
\begin{equation}
    \hat{U}_I = e^{- \ii \lambda \hat{\mu} \hat{\phi}(\Lambda)}.
\end{equation}
Assuming that the detector starts in an uncorrelated state with the field $\hat{\rho}_0 = \hat{\rho}_{\tc{d},0} \otimes \hat{\rho}_{\omega}$, the final state of the detector after tracing over the field's degrees of freedom can be written as~\cite{Landulfo,analytical}
\begin{align}
    \hat{\rho}_\tc{d} &= \tr_{\phi}(\hat{U}\hat{\rho}_0\hat{U}^{\dagger}) \label{eq:nogap}\\
    &= e^{- \xi}\cosh(\xi)\hat{\rho}_{\tc{d},0} + e^{-\xi}\sinh(\xi)\hat{\mu}\hat{\rho}_{\tc{d},0}\hat{\mu},\nonumber
\end{align}
where $\xi = \lambda^2 W(\Lambda,\Lambda) = \lambda^2 \langle\hat{\phi}(\Lambda)^2 \rangle_\omega$.  That is, the final state of a gapless detector is entirely determined by the expected value of the squared field amplitude. In particular, using Eq.~\eqref{eq:main} in our expression for the final state of the detector in~\eqref{eq:nogap}, we can also find the leading order corrections to the final state of the detector due to curvature in a region of size $\ell$, namely
\begin{align}
    \hat{\rho}_\tc{d} = e^{- \xi_0}&\cosh(\xi_0)\hat{\rho}_{\tc{d},0} + e^{-\xi_0}\sinh(\xi_0)\hat{\mu}\hat{\rho}_{\tc{d},0}\hat{\mu}
    \\& + \lambda^2 \mathcal{R}(\Lambda)(\hat{\rho}_{\tc{d},0} + \hat{\mu} \hat{\rho}_{\tc{d},0}\hat{\mu}) + \mathcal{O}(\ell^3),\nonumber
\end{align}
where
\begin{equation}
    \mathcal{R}(\Lambda) = -\frac{4 \pi^2}{3}R_{abcd}\mathcal{L}^{abcd}-\frac{ R_{a b}}{12} \mathcal{L}^{a b}  + \frac{1}{12} R \mathcal{P}_{\ln} +\omega_0 \mathcal{P}_{\Lambda},
\end{equation}
and the terms $\mathcal{L}^{abcd}$, $\mathcal{L}^{ab}$, $\mathcal{P}_{\ln}$ and $\mathcal{P}_{\Lambda}$ are defined in Eq.~\eqref{Ls}. We then see that the corrections for the expected value $\langle\hat{\phi}(\Lambda)^2\rangle_\omega$ in Section~\ref{sec:curvature} indeed yield the leading order effect of curvature experienced by local probes\footnote{Also notice that by replacing $\Lambda(\mf x)\Lambda(\mf x') \mapsto \Lambda^-(\mf x)\Lambda^+(\mf x)$ in the integrands of Eq.~\eqref{Ls}, we obtain the leading order curvature corrections for $\omega(\Lambda^-(\mf x)\Lambda^+(\mf x'))$, and consequently, the leading order correction to the excitation probability of a gapped particle detector in Eq.~\eqref{eq:L}.}. {It is worth noting that in~\cite{drg1,drg2,drg3} related results for the effect of curvature in pointlike detectors were computed in different scenarios.}



Finally, we briefly overview a special case of coupling of gapless detectors: the limit of ultra rapid coupling (delta coupling)~\cite{deltaCoupled,nogo,ahmed,ahmedUV,ericksonNonPertDelta}. A delta coupled detector couples to the field instantaneously in a spatial slice. It can be obtained as a particular case of a gapless detector by picking a spacetime smearing function of the form
\begin{equation}\label{eq:smearingDeltaCoupling}
    \Lambda(\mf x) = \delta(\tau - \tau_0) f(\bm \xi),
\end{equation}
where $\tau$ again denotes the Fermi normal coordinate time coordinate of the trajectory $\mf z(\tau)$ and $\bm \xi$ denote the spatial Fermi normal coordinates. This type of $\Lambda(\mf x)$ gives rise to a smeared field operator of the form
\begin{equation}
    \hat{Y}(f) = \int \dd^3 \bm \xi \sqrt{-\bar{g}} f(\bm \xi) \hat{\phi}(\tau_0,\bm \xi),
\end{equation}
where $\bar{g}$ denotes the metric determinant in Fermi normal coordinates. The operator $\hat{Y}(f)$ is entirely localized in the surface $\tau = \tau_0$. Albeit useful for describing fast interactions with a quantum field, formally, the limit of ultra rapid coupling is unphysical, given that it prescribes an infinitely strong interaction that is switched on instantaneously.

We remark that our expansions of Section~\ref{sec:curvature} cannot be directly applied to the case of delta coupled detectors, as it is non trivial to write Eq.~\eqref{eq:smearingDeltaCoupling} in Riemann normal coordinates. Indeed, in~\cite{ahmed} a similar expansion to that of Section~\eqref{sec:curvature} was performed\footnote{However, the expansion of Eq.~\eqref{eq:Matheus} was not taken into account in~\cite{ahmed}.}, where the leading order effect of curvature in the excitation probability of delta coupled detectors was obtained using Fermi normal coordinates. The expansions of~\cite{ahmed} include corrections to the excitation probability that depend on the details of the trajectory $\mf z(\tau)$, such as its acceleration and the shape of the local rest surfaces associated with it. These additional dependencies mainly arise from the effective redshift factor relating the expression for $\sqrt{-g}$ in Riemann and $\sqrt{-\bar{g}}$ in Fermi normal coordinates.

\section{Conclusions}\label{sec:conclusion}

In this work, we have investigated the leading-order effects of spacetime curvature on the expected value of localized quantum field observables, focusing on the squared field amplitude of a massless scalar field, which fully characterizes quasi-free states. By using Riemann normal coordinates and the Hadamard condition, we computed curvature corrections to the expected value of these localized observables. Our approach relies on the fact that the algebraic formulation of QFT rigorously defines localized field observables through spacetime smearing functions. These functions have an operational interpretation, corresponding to the localization of physical probes used to implement local operations in the context of quantum field theory. 

Looking ahead, our methods can be extended to more general observables and field theories, such as vector or fermionic fields. Moreover, in principle, the corrections we have derived could be done to higher orders in curvature. Ultimately, the expansion in curvature of the Wightman function of Hadamard states is a useful approach for studying the effect of curvature in local operations in quantum field theory.

Our results also provide the leading order correction to the measurement outcomes of localized probes modelled by Unruh-DeWitt detectors. In particular, for gapless detectors, we explicitly derived the leading-order curvature-induced changes to the detector’s final state, showing how such probes can directly measure spacetime-induced corrections to field fluctuations.

On the other hand, one can use these localized probes to measure field observables, and then infer the geometry of spacetime from these statistics. This approach could enable a type of quantum field tomography, where one recovers information about the geometry of spacetime by analyzing the quantum fluctuations measured by localized probes. Indeed, this concept has been explored in~\cite{ahmed}, and in~\cite{achim,achim2,geometry} it has been suggested that this approach could be used to define an effective spacetime geometry in scales beyond those of General Relativity.

\section*{Acknowledgements}
 MHZ thanks Profs. Eduardo Mart\'n-Mart\'inez and Achim Kempf for funding through their NSERC Discovery grants. TRP acknowledges support from the Natural Sciences and Engineering Research Council of Canada (NSERC) via the Vanier Canada Graduate Scholarship. Research at Perimeter Institute is supported in part by the Government of Canada through the Department of Innovation, Science and Industry Canada and by the Province of Ontario through the Ministry of Colleges and Universities. Perimeter Institute and the University of Waterloo are situated on the Haldimand Tract, land that was promised to the Haudenosaunee of the Six Nations of the Grand River, and is within the territory of the Neutral, Anishinaabe, and Haudenosaunee people.

\appendix

\section{Explicit computation of the coefficients in the short distance limit}\label{app:L-terms}

In this appendix, we explicitly evaluate the coefficients $\mathcal{L}^{ab}$ and $\mathcal{L}^{abcd}$ from Eqs. \eqref{Ls} in the short distance limit. To this end, we shall work with a Gaussian spacetime smearing prescribed in RNC, namely
\begin{equation}\label{eq:LambdaA1}
    \Lambda(\mf x) = \Lambda(t, \bm x)  = \frac{e^{-\frac{-t^2}{2 T^2}}}{\sqrt{2\pi}T}\frac{e^{-\frac{|\bm x|^2}{2 \sigma^2}}}{(2\pi \sigma^2)^{3/2}},
\end{equation}
where we employed the notation $x^{a} = (t, \bm x)$ for the RNC system centered at $\mf z$. For this specific choice of spacetime smearing, the term $\langle \hat{\phi}_{0}(\Lambda)^2 \rangle$ was already computed in Eq. \eqref{eq:nice}. To evaluate the term $\mathcal{L}^{ab}$, observe that the integral defining this term can be split into the sum of four integrals of the form
\begin{equation}
     \int\dd^4 \mf x \dd^4 \mf x' \Lambda^{\text{eff}}_{1}(\mf x) \Lambda^{\text{eff}}_{2}(\mf x') W_0(\mf x,\mf x'),
\end{equation}
where, in each one of the four cases, the effective smearings $\Lambda^{\text{eff}}_{1}(\mf x)$ and $\Lambda^{\text{eff}}_{2}(\mf x')$ incorporate the factors $x^{a}x^{b}$, $x^{a}{x'}^{b}$, ${x'}^{a}x^{b}$, and  ${x'}^{a}{x'}^{b}$, respectively. Thus, using the definition of the Fourier transform, Eq. \eqref{eq:fourier_ladrao}, one can reduce the evaluation of $\mathcal{L}^{ab}$ to a sum of four integrals of the form,
\begin{equation}
    \frac{1}{(2\pi)^3}\int \frac{\dd^3 \bm k}{2|\bm k|} \tilde{\Lambda}^{\text{eff}}_{1}(|\bm k|,\bm k)\tilde{\Lambda}^{\text{eff}}_{2}(|\bm k|, \bm k)^{*}.
    \label{eq:effective_lambdas}
\end{equation}
Then, with straightforward calculations, we obtain
\begin{equation}
    \mathcal{L}^{00} = \frac{T^2}{4 \pi^2(T^2 + \sigma^2)},
\end{equation}
and
\begin{equation}
    \mathcal{L}^{ij} = \frac{\sigma^2\delta^{ij}}{4 \pi^2(T^2 + \sigma^2)},
\end{equation}
with $i,j=1,2,3$. The $\mathcal{L}^{0j}$ components vanish.

Next, we evaluate the terms $\mathcal{L}^{abcd}$. First observe that
\begin{equation}
    \partial^{a}W_{0}(\mf x, 
    \mf x') = -8 \pi^2 W^{2}_{0}(\mf x, \mf x')(x - x')^{a}.
\end{equation}
Using this fact, the term $\mathcal{L}^{abcd}$ can be cast into
\begin{align}
    \mathcal{L}^{abcd}  =& \int\dd^4 \mf x \dd^4 \mf x' \Lambda(\mf x) \Lambda(\mf x') W_0^2(\mf x,\mf x')(x-x')^{a}x^{d}{x'}^{b}{x'}^{c} \nonumber \\
    & +\int\dd^4 \mf x \dd^4 \mf x' \Lambda(\mf x) \Lambda(\mf x') W_0^2(\mf x,\mf x'){x'}^{a}x^{d}{x'}^{b}{x'}^{c}.
\end{align}
Now, the last term will vanish when contracted with $R_{abcd}$ due to the symmetries of the Riemann tensor. Thus, for our purposes, we only need to consider
\begin{align}
    \mathcal{\tilde{L}}^{abcd} & = \int\dd^4 \mf x \dd^4 \mf x' \Lambda(\mf x) \Lambda(\mf x') W_0^2(\mf x,\mf x')(x-x')^{a}x^{d}{x'}^{b}{x'}^{c} \nonumber \\
    & = -\frac{1}{8 \pi^2}\int\dd^4 \mf x \dd^4 \mf x' \Lambda(\mf x) \Lambda(\mf x') \partial^{a}W_{0}(\mf x, \mf x')x^{d}{x'}^{b}{x'}^{c}.
\end{align}
Performing integration by parts, we can write
\begin{equation}
    \mathcal{\tilde{L}}^{abcd}= \frac{1}{8\pi^2 }\left(\mathcal{B}^{abcd} + \delta^{ad}\mathcal{A}^{bc}\right),
\end{equation}
where
\begin{equation}
    \mathcal{B}^{abcd} = \int\dd^4 \mf x \dd^4 \mf x' \partial^{a}\Lambda(\mf x) \Lambda(\mf x') W_0(\mf x,\mf x'){x'}^{b}{x'}^{c}x^{d},
    \label{eq:B_abcd}
\end{equation}
and
\begin{equation}
    \mathcal{A}^{bc} = \int\dd^4 \mf x \dd^4 \mf x' \Lambda(\mf x) \Lambda(\mf x') W_0(\mf x,\mf x'){x'}^{b}{x'}^{c}.
    \label{eq:A_bc}
\end{equation}
Notice that equations Eq. \eqref{eq:B_abcd} and \eqref{eq:A_bc} can be evaluated using the same method we used to obtain the expression for $\mathcal{L}^{ab}$. That is, we can use Eq. \eqref{eq:effective_lambdas} with effective spacetime smearings that will incorporate extra terms of $x^{a}$ and ${x'}^{b}$ and, in the case of $\Lambda^{\text{eff}}_{1}(\mf x)$, might include the derivative smearing $\partial^{a}\Lambda(\mf x)$. The results for the non-zero $\mathcal{A}^{ab} $ terms are
\begin{align}
    \mathcal{A}^{00} &= \frac{T^2 \sigma^2}{8 \pi^2(T^2 + \sigma^2)^2}, \nonumber \\
    \mathcal{A}^{ij} &= \delta^{ij}\frac{3T^2 \sigma^2 + 2 \sigma^4}{24 \pi^2 (T^2 + \sigma^2)^2}.
\end{align}
As for the $\mathcal{B}^{abcd} $ terms, we only need to evaluate those whose contraction with the Riemann tensor does not vanish trivially. The relevant non-zero terms for the choice of $\Lambda(\mf x)$ in Eq.~\eqref{eq:LambdaA1} are the following:
\begin{align}
    \mathcal{B}^{0i0j} &= \delta^{ij}\frac{T^2 \sigma^4}{12 \pi^2\left(T^2+\sigma^2\right)^3}, \nonumber \\
    \mathcal{B}^{i00j} &=-\delta^{ij}\frac{T^2 \sigma^2(2 T^2 +\sigma^2)}{12 \pi^2\left(T^2+\sigma^2\right)^3}, \nonumber \\
    \mathcal{B}^{0ij0}&=\delta^{ij}\frac{\sigma^4\left(2 T^2+\sigma^2\right)}{12 \pi^2\left(T^2+\sigma^2\right)^3}, \nonumber \\
      \mathcal{B}^{ijkl} = &-\frac{\sigma^2\left[\delta^{il}\delta^{jk}(15T^4 + 20T^2\sigma^2 + 7 \sigma^4) + 2\sigma^4 \delta^{ik}\delta^{jl}\right]}{120 \pi^2 (T^2 + \sigma^2)^3}.
\end{align}
Thus, the effective coefficients $\tilde{\mathcal{L}}^{abcd}$ which are needed to compute the curvature corrections are given by
\begin{align}
    \mathcal{\tilde{L}}^{0i0j} &=   \delta^{ij}\frac{T^2 \sigma^4}{96 \pi^4\left(T^2+\sigma^2\right)^3}, \nonumber \\
    \mathcal{\tilde{L}}^{i00j} &=  \delta^{ij} \frac{T^2\sigma^2(\sigma^2 - T^2)}{192 \pi^4\left(T^2+\sigma^2\right)^3}, \nonumber \\
    \mathcal{\tilde{L}}^{0ij0} &= \delta^{ij}\frac{3 T^4 \sigma^2+9 T^2 \sigma^4+4 \sigma^6}{192 \pi^4\left(T^2+\sigma^2\right)^3}, \nonumber \\
     \mathcal{\tilde{L}}^{ijkl}  &=\frac{\delta^{il}\delta^{jk}(5T^2 \sigma^4 + 3 \sigma^6) - 2 \delta^{ik}\delta^{jl}\sigma^{6}}{960 \pi^4 (T^2 + \sigma^2)^3}.
\end{align} 
Using these results, we can finally evaluate the corrections in Eq. \eqref{eq:main} due to the Riemann and Ricci tensors. The first correction is
\begin{equation}
    -\frac{1}{12}R_{ab}\mathcal{L}^{ab} = -\frac{T^2R_{00} + \sigma^2\sum_{i=1}^{3}R_{ii}}{48 \pi^2(T^2 + \sigma^2)},
\end{equation}
whereas the coupling with the Riemann tensor yields 
\begin{align}
    -\frac{4 \pi^2}{3}R_{abcd}\mathcal{L}^{abcd} =& -\frac{4 \pi^2}{3}R_{abcd}\mathcal{\tilde{L}}^{abcd}  \\ =&-\frac{4 \pi^2}{3}R_{0i0j}(\mathcal{\tilde{L}}^{0i0j} - \mathcal{\tilde{L}}^{i00j} - \mathcal{\tilde{L}}^{0ij0}) \nonumber\\ 
   &\:\:\:\:-\frac{4 \pi^2}{3} R_{ijij}(\mathcal{\tilde{L}}^{ijij} - \mathcal{\tilde{L}}^{ijji}) \nonumber\\
   = &\,\frac{\sigma^2\left(T^4+4 T^2 \sigma^2+2 \sigma^4\right)}{72 \pi^2\left(T^2+\sigma^2\right)^3}\sum_{i=1}^{3}R_{0i0i} \nonumber \\
   &\:\:\:\:+
\frac{\sigma^4}{144 \pi^2\left(T^2+\sigma^2\right)^2}\sum_{i,j=1}^{3}R_{ijij}\nonumber
\end{align}
Now, recall that in Eq. \eqref{eq:main} the components $R_{ab}$ and $R_{abcd}$ are expressed in RNC and evaluated at the origin $\mf z$. Thus, the Minkowski metric is used to raise and lower indices, and the following relations hold:
\begin{equation}
    \sum_{i=1}^{3}R_{0i0i} = R_{00}
\end{equation}
and
\begin{equation}
    \sum_{i,j=1}^{3}R_{ijij} = \sum_{i=1}^{3}R_{ii} +R_{00} = R^{i}_{i}+R_{00}.
\end{equation}
Therefore, for a Gaussian spacetime smearing prescribed in RNC, the corrections to Eq. \eqref{eq:main} due to the Ricci and Riemann tensors take the form
\begin{align}
  &-\frac{1}{12}R_{ab}\mathcal{L}^{ab}  -\frac{4 \pi^2}{3}R_{abcd}\mathcal{L}^{abcd} \nonumber \\
  & = \frac{-3 T^6-4 T^4 \sigma^2+6 T^2 \sigma^4+5 \sigma^6}{144 \pi^2\left(T^2+\sigma^2\right)^3}R_{00}  \nonumber \\
  & -\frac{3T^2 \sigma^2+2 \sigma^4}{144 \pi^2\left(T^2+\sigma^2\right)^2}R^{i}_{i}.
\end{align}
In particular, by setting $\sigma = T$ one obtains
\begin{align}
       -\frac{4 \pi^2}{3}R_{abcd}&\mathcal{L}^{abcd}-\frac{1}{12} R_{a b} \mathcal{L}^{a b}  \\&= \frac{2 R_ {00}-5 R^{i}_{i}}{576 \pi^2} = -\frac{5R + 3R_{00}}{576 \pi^2}.\nonumber
\end{align}

\section{Expansion of the Synge's world function in Riemann normal coordinates}\label{app:sing}

In this appendix, we expand the Synge's world function in a RNC system centered at $\mf z$ to establish the results of Eqs. \eqref{eq:camelsing}. For the sake of completeness, we start by recalling the definition of the Synge's world function:
\begin{equation}
    \sigma(\mf x, \mf x') =\frac{1}{2}(\lambda_2 -\lambda_1)\int_{\lambda_1}^{\lambda_2}{\dd \lambda \  g_{\mu \nu}(\xi(\lambda))v^{\mu}(\lambda)v^{\nu}(\lambda)},
    \label{eq:synge_roots}
\end{equation}
where $\xi(\lambda)$ is a timelike geodesic connecting the points $\mf x = \xi(\lambda_{1})$ and $\mf x' = \xi(\lambda_{2})$, and $v = \dot{\xi}(\lambda)$ is the vector field tangent to the trajectory.

We use Latin indices $a, b, c, \ldots$ to denote the components of tensors in the RNC system centered at $\mf z$. The expansion of the metric tensor up to second order in curvature then reads \cite{poisson}
\begin{equation}
    g_{a b}(\xi(\lambda)) \approx \eta_{a b}(\mf z)+\frac{1}{3} R_{a c d b}(\mf z) \xi(\lambda)^c \xi(\lambda)^d. 
    \label{eq:metric_tensor_RNC}
\end{equation}
Substituting Eq.\eqref{eq:metric_tensor_RNC} in Eq. \eqref{eq:synge_roots}, it follows that
\begin{equation}
    \sigma(\mf x, \mf x') \approx \frac{1}{2}(\lambda_2-\lambda_1)^2+\frac{1}{6} R_{a c d b}(\mf z) \int_{\lambda_1}^{\lambda_2} \dd \lambda \  \xi^c \xi^d v^a v^b.
    \label{eq:sing_the_song}
\end{equation}
Next, we consider the expansion of the geodesic $\xi(\lambda)$ using the RNC of the fixed point $\mf x$, namely
\begin{equation}
    \xi^{a}(\lambda) = x^{a}+ \lambda v_{0}^{a} + \frac{1}{2}A^{a}\lambda^2 + \mathcal{O}(\lambda^3),
\end{equation}
where $v_{0} = \dot{\xi}(0)$. To determine the coefficients $A^{a}$, we use the approximation
\begin{equation}
    \Gamma^{a}_{\ b c} \approx \Gamma^{a}_{\ b c}(\mf z) + \partial_{d}\Gamma^{a}_{\ b c}(\mf z)x^{d} = \partial_{d}\Gamma^{a}_{\ b c}(\mf z)x^{d}
\end{equation}
and the expansion for $\xi(\lambda)$ into the geodesic equation,
\begin{equation}
    \frac{d \xi^{a}}{d \lambda^2} + \Gamma^{a}_{\ b c}\dot{\xi}^{b}\dot{\xi}^{c} = 0.
\end{equation}
Next, we solve it perturbatively in the parameter $\lambda$. In this way, one obtains
\begin{equation}
    A^{a} = -\partial_{d}\Gamma^{a}_{\ bc}(\mf z)x^{d}v_{0}^{b}v^{c}_{0}.
\end{equation}
Now, because the point $\mf z$ is the centre of the RNC system, we can write
\begin{equation}
    R^{a}_{ \ b c d}(\mf z)=\partial_c \Gamma_{b d}^a(\mf z)-\partial_d \Gamma_{b c}^a(\mf z).
\end{equation}
Combining this expression with the symmetries of the Riemann tensor, we get
\begin{equation}
    \partial_b \Gamma^{a} _{\ c d}(\mf z)=-\frac{1}{3}(R^{a}_{ \ c d b}(\mf z)+R^{a}_{\ d c b}(\mf z)).
\end{equation}
Then, the expansion of the geodesic $\xi(\lambda)$ can be rewritten as
\begin{equation}
    \xi^{a}(\lambda) = x^a+\lambda v_0^a+\frac{1}{3} v_0^b v_0^c x^d R^{a}_{\ b c d}(\mf z)\lambda^2 + \mathcal{O}(\lambda^3).
    \label{eq:another_day_another_expansion}
\end{equation}
At this point, recall that our goal is to write an expansion for $\sigma(\mf x, \mf x')$ keeping only terms to first order in the curvature. Thus, in Eq. \eqref{eq:another_day_another_expansion}, we can effectively use the approximations $ \xi^{a} \approx x^{a} + \lambda v_{0}^{a}$ and $v^{a} \approx v^{a}_{0}$. Then, by plugging the resulting expansion into Eq.\eqref{eq:sing_the_song}, we have 
\begin{equation}
    \sigma(\mf x, \mf x') \approx \frac{1}{2}(\lambda_2-\lambda_1)^2+\frac{1}{6}(\lambda_2-\lambda_1)^2 R_{acdb} x^c x^d v_0^a v_0^b.
    \label{eq:sigma_expan}
\end{equation}
Finally, notice that
\begin{equation}
    (\lambda_1 - \lambda_2)^2 = \eta_{ab}(x^{\prime} - x)^{a}(x^{\prime} - x)^{b}.
\end{equation}
Moreover, we can approximate $(\lambda_{2} - \lambda_{1}) v^{a}_{0} \approx {x'}^{a} - x^{a}$, where the curvature terms are being neglected in order to avoid second order contributions in Eq. \eqref{eq:sigma_expan}. Therefore, our final expansion for $\sigma(\mf x, \mf x')$ to first order in the curvature reads
\begin{equation}
    \sigma(\mf x, \mf x') \approx \frac{1}{2}\eta_{ab}(x-x')^a(x-x')^b + \frac{1}{6} R_{acbd}(\mf z) x^a x^b x'{}^c x'{}^d.
    \label{eq:jacob}
\end{equation}
The expansion for the derivatives, $\sigma_{a} = \partial_{a}\sigma$ and \mbox{$\sigma_{a'} = \partial_{a'} \sigma$}, are obtained straightforwardly. Indeed, by taking the partial derivative of Eq.\eqref{eq:jacob} with respect to the coordinates $x^{a}$, we have
\begin{equation}
    \sigma_{a}(\mf x, \mf x')\approx (x-x')_a + \frac{1}{3} R_{acbd}{}(\mf z) x^b x'{}^c x'{}^d.
\end{equation}
Similarly,
\begin{equation}
    \sigma_{c'}(\mf x, \mf x') \approx (x-x')_c + \frac{1}{3} R_{acbd}(\mf z) x^a x^b x'{}^d.
\end{equation}

\bibliography{references}
\end{document}